\documentclass[12pt, epsfig]{article}
\usepackage{epsfig}

\begin{document}


\newcommand{\beq}{\begin{equation}}
\newcommand{\eeq}{\end{equation}}
\newcommand{\bea}{\begin{eqnarray}}
\newcommand{\eea}{\end{eqnarray}}
\newcommand{\beqn}{\begin{eqnarray}}
\newcommand{\eeqn}{\end{eqnarray}}
\newcommand{\beas}{\begin{eqnarray*}}
\newcommand{\eeas}{\end{eqnarray*}}
\newcommand{\defi}{\stackrel{\rm def}{=}}
\newcommand{\non}{\nonumber}
\newcommand{\bquo}{\begin{quote}}
\newcommand{\enqu}{\end{quote}}
\newcommand{\p}{\partial}


\def\de{\partial}
\def\Tr{ \hbox{\rm Tr}}
\def\const{\hbox {\rm const.}}
\def\o{\over}
\def\im{\hbox{\rm Im}}
\def\re{\hbox{\rm Re}}
\def\bra{\langle}\def\ket{\rangle}
\def\Arg{\hbox {\rm Arg}}
\def\Re{\hbox {\rm Re}}
\def\Im{\hbox {\rm Im}}
\def\diag{\hbox{\rm diag}}

\def\stroke{\vrule height8pt width0.4pt depth-0.1pt}
\def\topfleck{\vrule height8pt width0.5pt depth-5.9pt}
\def\botfleck{\vrule height2pt width0.5pt depth0.1pt}
\def\Zmath{\vcenter{\hbox{\numbers\rlap{\rlap{Z}\kern 0.8pt\topfleck}\kern
2.2pt\rlap Z\kern 6pt\botfleck\kern 1pt}}}
\def\Qmath{\vcenter{\hbox{\upright\rlap{\rlap{Q}\kern
3.8pt\stroke}\phantom{Q}}}}
\def\Nmath{\vcenter{\hbox{\upright\rlap{I}\kern 1.7pt N}}}
\def\Cmath{\vcenter{\hbox{\upright\rlap{\rlap{C}\kern
3.8pt\stroke}\phantom{C}}}}
\def\Rmath{\vcenter{\hbox{\upright\rlap{I}\kern 1.7pt R}}}
\def\Z{\ifmmode\Zmath\else$\Zmath$\fi}
\def\Q{\ifmmode\Qmath\else$\Qmath$\fi}
\def\N{\ifmmode\Nmath\else$\Nmath$\fi}
\def\C{\ifmmode\Cmath\else$\Cmath$\fi}
\def\R{\ifmmode\Rmath\else$\Rmath$\fi}


\def\QATOPD#1#2#3#4{{#3 \atopwithdelims#1#2 #4}}
\def\stackunder#1#2{\mathrel{\mathop{#2}\limits_{#1}}}
\def\stackreb#1#2{\mathrel{\mathop{#2}\limits_{#1}}}
\def\Tr{{\rm Tr}}
\def\res{{\rm res}}
\def\Bf#1{\mbox{\boldmath $#1$}}
\def\balpha{{\Bf\alpha}}
\def\bbeta{{\Bf\beta}}
\def\bgamma{{\Bf\gamma}}
\def\bnu{{\Bf\nu}}
\def\bmu{{\Bf\mu}}
\def\bphi{{\Bf\phi}}
\def\bPhi{{\Bf\Phi}}
\def\bomega{{\Bf\omega}}
\def\blambda{{\Bf\lambda}}
\def\brho{{\Bf\rho}}
\def\bsigma{{\bfit\sigma}}
\def\bxi{{\Bf\xi}}
\def\bbeta{{\Bf\eta}}
\def\d{\partial}
\def\der#1#2{\frac{\d{#1}}{\d{#2}}}
\def\Im{{\rm Im}}
\def\Re{{\rm Re}}
\def\rank{{\rm rank}}
\def\diag{{\rm diag}}
\def\2{{1\over 2}}
\def\ntwo{${\cal N}=2\;$}
\def\4N{${\cal N}=4$}
\def\none{${\cal N}=1\;$}
\def\x{\stackrel{\otimes}{,}}
\def\beq{\begin{equation}}
\def\eeq{\end{equation}}
\def\ba{\beq\new\begin{array}{c}}
\def\ea{\end{array}\eeq}
\def\be{\ba}
\def\ee{\ea}
\def\stackreb#1#2{\mathrel{\mathop{#2}\limits_{#1}}}

\def\baselinestretch{1.0}

\begin{titlepage}

\begin{flushright}
FTPI-MINN-05/09, UMN-TH-2351/05\\
ITEP-TH-25/05\\
April 18, 2005
\end{flushright}

\vspace{1cm}

\begin{center}

{\Large \bf   Studying Boojums in \boldmath{${\cal N}=2$}
 \\[2mm] Theory with Walls and Vortices}
\end{center}

\vspace{0.5cm}

\begin{center}
{\bf R. Auzzi,$^{a}$  M.~Shifman,$^{a}$ and \bf A.~Yung$^{a,b,c}$}
\end {center}
\begin{center}

$^a${\it  William I. Fine Theoretical Physics Institute,
University of Minnesota,
Minneapolis, MN 55455, USA}\\
$^{b}${\it Petersburg Nuclear Physics Institute, Gatchina, St. Petersburg
188300, Russia}\\
$^c${\it Institute of Theoretical and Experimental Physics, Moscow
117259, Russia}
\end{center}

\vspace{3mm}

\begin{abstract}

We study 1/2 BPS domain walls,  1/2 BPS flux tubes (strings)
and their 1/4 BPS junctions. We consider the
simplest example of \ntwo Abelian gauge
theory with two charged matter hypermultiplets which contains all
of the above-listed extended objects. In particular, we focus on
string-wall junctions (boojums) and calculate their energy. It turns out
to be logarithmically divergent in the infrared domain. We compute this energy
first in the (2+1)-dimensional effective theory on the domain wall
and then, as a check, obtain the same result from the point of view
of (3+1)-dimensional bulk theory. Next, we study interactions of
boojums considering all possible geometries of string-wall junctions
and directions of the string magnetic fluxes.

\end{abstract}

\end{titlepage}

\section{Introduction}
\label{intro}

We are witnessing remarkable advances in the
{\em field-theoretic} studies of branes (domain walls) and strings
(flux tubes) \cite{1,2,3,sy1,sy2,4,5} with the purpose of revealing and exploring parallels
between supersymmetric field theories and string/D-brane theory
(for an exhaustive review covering also the issue of confined monopoles
see \cite{sakai-tong}).
The unfolding of this program led us well beyond
initial anticipations \cite{2,sy1,sy2}.

Supersymmetric  gauge theories in $(3+1)$ dimensions
in the Higgs phase
have  a rich spectrum of solitons, including domain walls, flux tubes
(vortices),  confined monopoles and all sorts of junctions.
The most convenient set-up exhibiting these solitons,
both ``elementary" and ``composite," which emerged in the recent years,
is provided by U$(N)$ Yang--Mills theory with
extended ${\cal N}=2$ supersymmetry (SUSY) and the
Fayet--Iliopoulos term for the $U(1)$ factor (see Refs.
\cite{HT,MY,vortici,tong-monopolo,
sy2,SY-vortici,HT2,inos,einos,monovortice}).  The matter sector includes
$N_f\geq N$ hypermultiplets.

Among other intriguing features revealed {\em en route},
let us mention just a few results: (i)
localization of an effective  U(1)
gauge theory   on an elementary domain wall; (ii)
demonstration that Abelian vortices can end on  walls and
act as massive magnetic sources
for the effective  U(1)  gauge theory localized on the wall;
(iii) localization of non-Abelian gauge fields on composite walls,
with non-Abelian flux tubes ending on them, etc.

Besides giving an extensive review of the subject,
Ref.~\cite{sakai-tong}
presents a new result regarding the wall-string junctions,
{\em boojums}\,\footnote{The word {\em boojum} comes from the Lewis Carroll's children's  book
{\em Hunting of the
Snark}. Apparently,  it is fun to hunt a snark, but if the snark
turns out to be a boojum, you are in trouble! Condensed matter
physicists adopted the name in the 1970s to describe solitonic
objects of the wall-string junction type in helium-3 and other systems.
Also: {\em boojum tree} (Idria  Columnaris) is a tree
endemic to  the  Baja California peninsula of Mexico.
The boojum  tree is one of the strangest plants imaginable. For most  of the year it is leafless and looks like a giant  upturned turnip. Its common name was coined by the plant  explorer Godfrey Sykes, who found it in 1922 and said  ``It must be a boojum!"  In saying this, Sykes was  referring
to Lewis Carrol's  book. The Spanish  common name for this tree is Cirio,  referring to its candle-like appearance. Supersymmetric wall-string junctions
were referred to as {\em boojums}
in Ref.~\cite{sakai-tong}.
}.
Our task is to further elaborate on this issue.
In this paper we discuss  aspects of the binding energy
of flux tubes ending on walls in the
``minimal" setting of  ${\cal N}=2$ SQED with two flavor
hypermultiplets. We also dwell on multi-boojums.
We find that the negative finite binding energy
observed in \cite{sakai-tong} is in fact ill-defined and
shadowed by certain logarithmic contributions in single
boojums, while it remains well-defined for certain multi-boojums.
In addition we consider forces between the  flux tubes
in the multi-boojum configurations.

The organization of the paper is as follows.
The results of Ref.~\cite{sy1} relevant for the subsequent consideration
are summarized in Sect.~\ref{bwv}. Section~\ref{vwj} presents
a calculation of energy of the wall-vortex system.
In Sect.~\ref{tvewi} various two-boojum  configurations, in which two vortices
end on the same wall, are discussed. In Sect.~\ref{bsest} we formulate 
a slightly modified approach  to define a boojum self-energy, while 
Sect.~\ref{conclu}  contains our conclusions.

\section{The bulk, the wall \& the vortex}
\label{bwv}

The bulk theory is
$\mathcal{N}=2$ SQED with 2 flavors and
a Fayet--Iliopoulos term. The bosonic part of the Lagrangian is
\beqn
&& S=\int d^4 x \left\{ \frac{1}{4 g^2} F_{\mu \nu}^2 +
\frac{1}{g^2} |\partial_\mu a|^2 +\bar{\nabla}_\mu \bar{q}_A \nabla_\mu q^A +
\bar{\nabla}_\mu \tilde{q}_A \nabla_\mu \bar{\tilde{q}}^A
\right.\nonumber\\[3mm]
 &&
 +
 \left.\frac{g^2}{8}(|q^A|^2-|\tilde{q}_A|^2-\xi)+\frac{g^2}{2} |\tilde{q}_A q^A|^2
+\frac{1}{2}(|q^A|^2+|\tilde{q}^A|^2)|a+\sqrt{2}m_A|^2 \right\}, \nonumber\\
\label{n2sqed}
\eeqn
where $$\nabla_\mu=\partial_\mu-\frac{i}{2}A_\mu$$ and
$$\bar{\nabla}_\mu=\partial_\mu+\frac{i}{2}A_\mu\,.$$
The index $A=1,2$ is the flavor index; the mass parameters $m_1,m_2$ are
assumed to be  real.
We will work in the limit
$$
\Delta m=m_1-m_2\gg g\sqrt{\xi} \,.
$$
There are two vacua in this theory: in the first vacuum
\beq
a=-\sqrt{2} m_1,\qquad q_1=\sqrt{\xi}, \qquad  q_2=0\,,
\label{fv}
\eeq
and in the second one
\beq
a=-\sqrt{2} m_2, \qquad q_1=0,  \qquad q_2=\sqrt{\xi}\,.
\label{sv}
\eeq

The vacuum expectation value (VEV) of
field $\tilde{q}$ vanishes in both  vacua.
Hereafter we will stick to the {\em ansatz} $\tilde{q}=0$.

\begin{figure}[h]
\epsfxsize=9cm
\centerline{\epsfbox{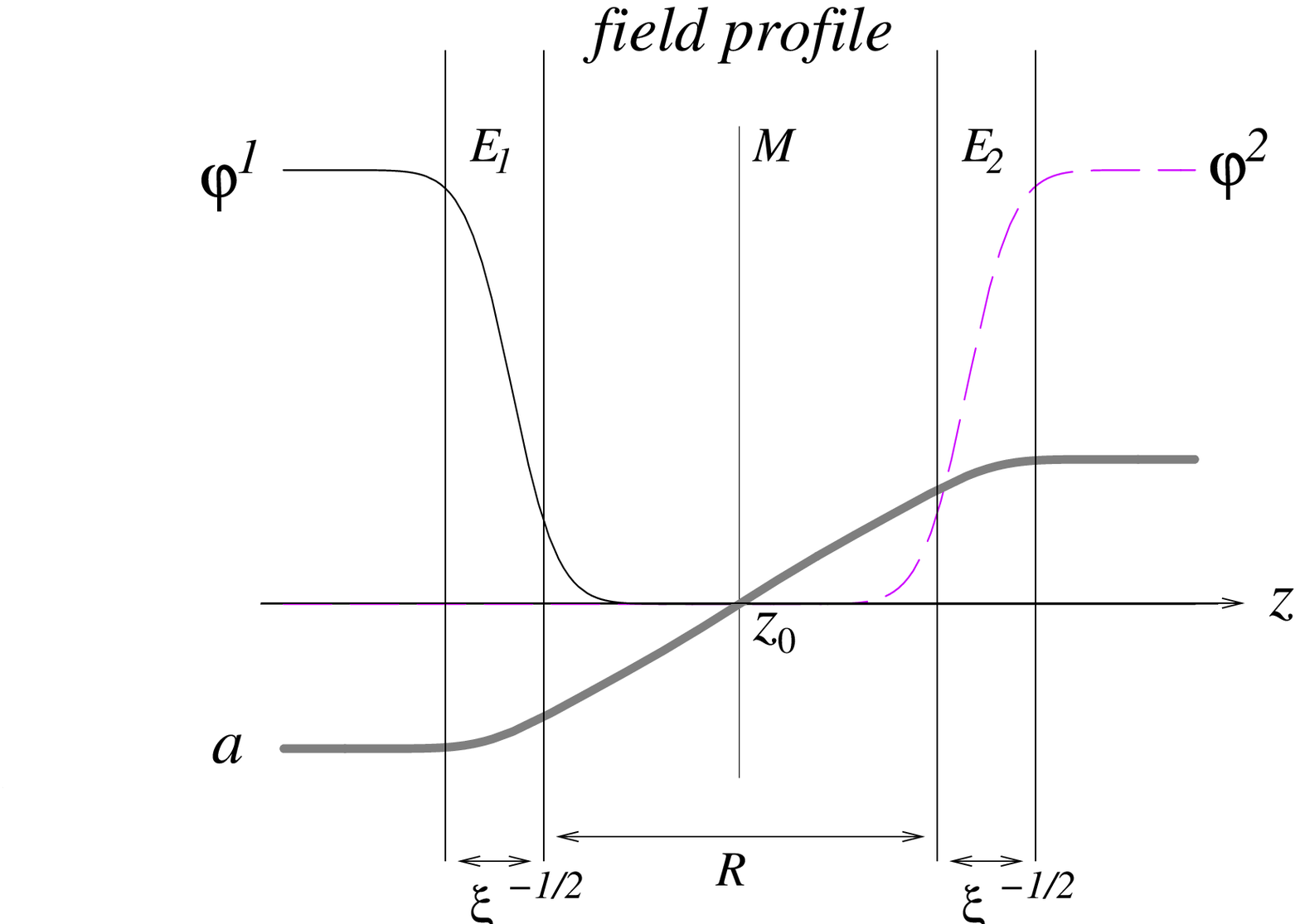}}
 \caption{\footnotesize Internal structure of the domain wall:
two edges (domains $E_{1,2}$) of the width $\sim \xi^{-1/2}$
are separated by a broad middle band (domain $M$) of the width $R
\sim \Delta m /(g^2\xi)$.}
\label{syfigthree}
\end{figure}

A BPS domain wall
interpolating between the two vacua of our bulk
theory was explicitly constructed in Ref. \cite{sy1}.
Assuming that all fields depend only on the coordinate
$z=x_3$, it is possible to write the energy in the Bogomolny form \cite{Bogomolny},
\beqn
  \int d^3 x \, \mathcal{H} &=&
 \int dx_3 \left\{ \left|\nabla_3 q^A \pm \frac{1}{\sqrt{2}}q^A
 (a+\sqrt{2}m_A)\right|^2\right.
 \nonumber\\[3mm]
&+&\left. \left|\frac{1}{g}\partial_3 a \pm \frac{g}{2 \sqrt{2}}
\left(|q^a|^2-\xi\right)\right|^2
\pm \frac{1}{\sqrt{2}} \xi \partial_3 a\right\}.
\label{bogfw}
\eeqn
Putting the first two terms above to zero gives us the BPS equations
for the wall. Assuming that $\Delta m >0$ we choose the upper sign in
(\ref{bogfw}).
The  tension is given by the total derivative term,
(the last one in Eq.~(\ref{bogfw}))
which can be identified as the $(1,0)$ central charge
of the supersymmetry algebra,
\beq
T_{\rm w}=\xi \, \Delta m\,.
\label{wten}
\eeq
The wall solution has a three-layer structure
(see Fig.~\ref{syfigthree}):
in the two outer layers (which have width
${O}(({g \sqrt{\xi}})^{-1}$)) the squark fields drop
to zero exponentially;  in the inner layer
the field $a$ interpolates between its two vacuum values.
The  thickness of this inner layer is given by
\beq
R=\frac{4 \Delta m}{g^2 \xi}.
\label{til}
\eeq

This wall is an $1/2$ BPS solution of the four-dimensional
action. In other words, the soliton breaks 4 of the 8 supersymmetry
generators of $\mathcal{N}=2$ bulk theory.
The moduli space
is described by two bosonic coordinates:
one of these coordinates is associated with the wall translation;
the other one is an  U(1)  compact parameter.
It is possible to promote these   moduli
to fields depending on the wall coordinates $x_n$ ($n=0,1,2$).
A world-volume theory for the moduli fields on the wall
can be written. The bosonic part of the world-volume
action is
\beq
S_{2+1}=\int d^3 x \left( \frac{T_{\rm w}}{2}
(\partial_n \zeta)^2+\frac{1}{4 e^2_{2+1}} ( F_{mn}^{(2+1)} )^2   \right)
,
\label{3wwa}
\eeq
where $\zeta$ describes the translational mode while
$e^2_{2+1}$ is the coupling constant of the
effective  U(1)  on the wall,
\beq
e^2_{2+1}=4 \pi^2 \frac{\xi}{\Delta m}\,.
\label{3cc}
\eeq
The fermion content of
the world-volume theory
is given by two three-dimensional Majorana spinors, as
is requited by ${\cal N}=2$ in three dimensions.
The full world-volume theory is
 U(1)  Yang--Mills theory in $(2+1)$ dimensions, with
four supercharges. The Lagrangian and the
corresponding superalgebra
can be obtained by reducing four-dimensional
$\mathcal{N}=1$ SQED (with no matter)
to three dimensions.

Our theory (\ref{n2sqed})  is an Abelian theory so it does not have
monopoles. However we can start from non-Abelian \ntwo theory with
the gauge group SU(2) broken down to U(1) theory (\ref{n2sqed})
by the condensation of the adjoint scalar field (whose third component
is $a$, see (\ref{fv}) and (\ref{sv})). This non-Abelian theory have 't
Hooft-Polyakov monopoles. In the low-energy theory (\ref{n2sqed}) they
become heavy external magnetic charges with mass of order of $m/g^2$.
As soon as electrically charged states condense in both vacua (\ref{fv})
and (\ref{sv}) these monopoles are in the confinement phase in both
vacua.

In fact,
in each of the two vacua of the bulk theory
magnetic charges are confined by the Abrikosov--Nielsen--Olesen
(ANO) strings.
In the vacuum in which $a=\sqrt{2} m_1$ we can use the
{\em ansatz} $q_2=0$ and
write the following BPS equations (see \cite{Bogomolny,VY}):
\beq
F_3^*-\frac{g^2}{2}(|q^1|^2-\xi)=0, \,\,\,(\nabla_1-i\nabla_2)q^1=0,
\label{fluxone}
\eeq
where
$$
F^{*}_i=1/2 \varepsilon_{ijk} F_{jk}\,,\qquad i,j,k=1,2,3\,.
$$
Analogous equations can be written for the flux tube
in the other vacuum, $a=\sqrt{2} m_2$.
These objects
are half-critical, much in the same way
as the domain walls above.

The magnetic flux of the minimal-winding flux tube
is $4 \pi$, while its    tension   is given by
the $(1/2,1/2)$ central charge \cite{GS},
\beq
T_{\rm s} =2 \pi \xi\,.
\label{tens}
\eeq
The thickness of the tube is of the order of
\beq
r_0 = (g^2\,\xi)^{-1/2}\,.
\label{tt}
\eeq

\section{The boojum (wall--flux-tube junction)}
\label{vwj}

Let us consider  a flux tube ending on a domain wall;
this configuration has been studied  in Refs.~\cite{sy1,sy2,inos,sakai-tong}.
Our primary goal is  examination of the binding energy.

The magnetic flux carried by the monopole is unconfined
inside the domain wall; so the vortex-wall junction behaves
as a charge in the Coulomb phase of the wall world-volume theory.
Let us consider a vortex oriented in the $z$ direction
ending on a wall oriented in the $x_1,\, x_2$ directions;
let the string extend at $z>0$ and let the magnetic
flux be oriented in the positive $z$ direction.
BPS first order equations can
be written for the composite soliton,
\beqn
&& F_1^*-iF_2^* + \sqrt{2}(\partial_1-i\partial_2)a=0,\nonumber\\[2mm]
&& F_3^*+\frac{g^2}{2} (|q^A|^2 - \xi)+\sqrt{2} \partial_3 a=0,\nonumber\\[2mm]
&&\nabla_3 q^A=-\frac{1}{\sqrt{2}}q^A (a+\sqrt{2}m_A), \nonumber\\[2mm]
&& (\nabla_1-i\nabla_2)q^A=0\,.
 \label{bpseq}
 \eeqn
These equations generalize both the $1/2$ BPS wall and the flux-tube equations
and can be used for determination of the general $1/4$ BPS boojum soliton,
(i.e. a flux tube ending on a wall).

\begin{figure}[h]
\epsfxsize=6cm
\centerline{\epsfbox{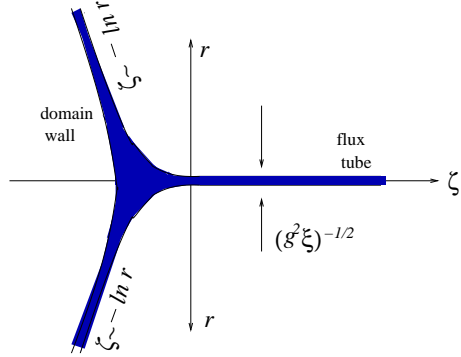}}
 \caption{\footnotesize  Bending of the wall due to the string-wall junction.
The flux tube extends to the right infinity. The wall
profile is logarithmic at transverse distances larger than
$g^{-1}\xi^{-1/2}$ from the string axis. At smaller distances
the adiabatic approximation fails.}
\label{syfigsix}
\end{figure}

At large distance $r$ from the string junction the wall is logarithmically
bent due to the fact that the vortex pulls it (see Fig.~\ref{syfigsix}),
\beq
z= -\frac{1}{\Delta m} \ln r+{\rm const}\, .
\label{shapeb}
\eeq

First of all, let us  estimate the binding energy
of the boojum from the standpoint   of
the world-volume theory.
At large distances $r$ from the string-wall junction
the fields in the (2+1) effective theory
are given by the following expressions:
\beq
F_{0i}^{2+1}=\frac{e^2_{2+1}}{2 \pi } \frac{x_i}{r^2} \,,\qquad
\zeta=-\frac{1}{\Delta m}\ln r +
{\rm const},
\label{ldr}
 \eeq
where $i=1,2$.
There are two contributions to the energy of this field
configuration.
The first contribution to the energy
localized at $r_0<r<r_f$ is due to
the gauge field,
\beqn
E^G_{(2+1)} &=& \int_{r_0}^{r_f}
\frac{1}{2 e^2_{2+1}} (F_{0i})^2\,  2 \pi r\, dr
\nonumber\\[4mm]
&=&\frac{\pi \xi}{\Delta m
}\int_{r_0}^{r_f}  \frac{ dr}{r}=\frac{\pi \xi}{\Delta m
} \ln \frac{r_f}{r_0}.
\label{gfcc}
\eeqn
The second contribution, due to the $\zeta$  field is,
\beqn
E^H_{(2+1)} &=& \int_{r_0}^{r_f}  \frac{T_{\rm w}}{2} (\partial_r \zeta)^2 \, 2 \pi r\,  dr
\nonumber\\[4mm]
&=&
\frac{\pi \xi}{\Delta m
}\ln \frac{r_f}{r_0}\,.
\label{zfcc}
\eeqn

\vspace{1mm}

These two contributions are both logarithmically divergent
at $r_f\rightarrow \infty$, this is an infrared (IR) divergency.
Its occurrence is an obvious
feature of the charged objects in $(2+1)$ dimensions due to the
fact that the fields vanish too slowly at  infinity to have
a finite energy.
There is no divergence in the ultraviolet (UV) domain, at   $r_0$.
In fact, the   value of $r_0$ is given by the thickness of the flux tube,
which is of the order of $r_0 \sim (g \sqrt{\xi})^{-1}$.

The above two contributions are equal (with the logarithmic
accuracy), even though their physical
interpretation is different. The total ``additional" energy is
\beq
E^{G+H}=\frac{2\pi \xi}{\Delta m
}\ln \frac{r_f}{r_0}\,.
\label{eto}
\eeq

\begin{figure}[h]
\epsfxsize=8cm
\centerline{\epsfbox{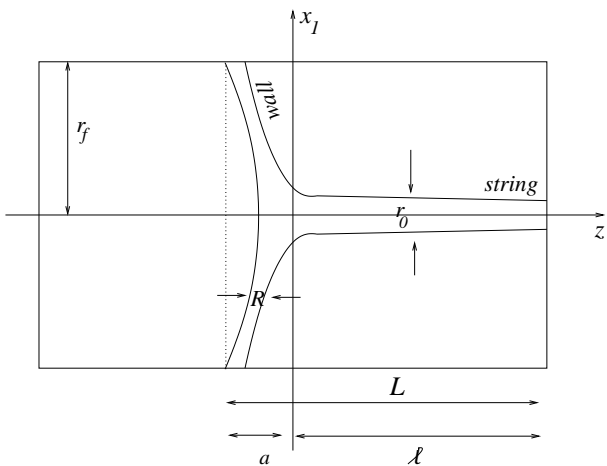}}
 \caption{\footnotesize Geometry of the boojum configuration}
\label{auone}
\end{figure}

The gauge field contribution
can be interpreted in the bulk theory as the energy carried by
the magnetic field.
The corresponding energy density is
\beq
\rho^{G} =\frac{1}{2 g^2}\,  \vec{B}^2
\eeq
inside the domain wall of
thickness $R$. This can be seen from the following
$(3+1)$ bulk
calculation:
\beq
E^G=\int_{r_0}^{r_f} \frac{1}{2 g^2}\,  |\vec{B}|^2 \, R
\, 2 \pi r \, dr=\frac{\pi \xi}{\Delta m
}\, \ln \frac{r_f}{r_0}\, .
\label{fbc}
\eeq
The $\zeta$  field contribution can be interpreted in $(3+1)$ dimensions as
due to the fact that the domain wall bends and, therefore,  its area $S$ changes,
$\Delta S =\pi (\Delta m)^{-2}\ln (r_f/r_0)$, and
\beqn
E^H &=& T_{\rm w}\,\Delta S
\nonumber\\[2mm]
&=& T_{\rm w} \int_{r_0}^{r_f} \left(\sqrt{1+\frac{1}{(\Delta m)^2 r^2 }}-1\right)
 2 \pi r \, dr\, \approx \frac{\pi \, \xi}{\Delta m}\, \ln \frac{r_f}{r_0}\,.
\label{cwa}
\eeqn
Equation
(\ref{cwa}) implies that the ``Higgs-induced" logarithmic term in
the energy of the boojum configuration (its geometry is depicted in
Fig.~\ref{auone}) can be absorbed
in the actual  area  of the curved wall,
\beq
S= \pi\,r_f^2 + \Delta S \,.
\label{manone}
\eeq
One then  measures the string length from the origin,
see the parameter $\ell$
in Fig.~\ref{auone}. This interpretation seems  natural.

Our composite soliton is quarter-critical. Therefore,  it is possible
to use the central charges approach to calculate its energy.
There are three types of central charges in $\mathcal{N}=2$
theories: the $(1,0)$ charge, which can be saturated by
domain walls; the $(1/2,1/2)$ charge, which can be saturated by  vortices;
the (${\cal N}=2$) Lorentz-scalar charge, which can be saturated by  monopoles.
When a BPS composite soliton such as a boojum  is treated, we have to consider all these
contributions to the total energy of the object that we are considering.
Assuming that BPS equations are satisfied, the energy of
the composite  soliton
is given by
\beq \int d^3 x \mathcal{H}=\int d^3 x \left\{ \frac{1}{\sqrt{2}} \xi \partial_3 a
+\xi \frac{F^{*}_3}{2} -\frac{1}{g^2} \sqrt{2}\,
\partial_\alpha\left(a F^{*}_\alpha\right)
\right\},
\label{velvet}
\eeq
where the three terms are respectively the $(1,0)$, the $(1/2,1/2)$ and the
Lorentz scalar  central charges (the latter is ${\cal N}=2$).
Let us consider the energy inside the domain
$x_1^2+x_2^2<r_f^2$
and of length $\ell$ in the $z$ direction. The distance $\ell $
is measured starting from the intersection of the tube
with the wall surface. As is clear from Fig.~\ref{auone},
\beq
L= a+\ell
\label{mantwo}
\eeq
where
\beq
a = \frac{1}{\Delta m}\, \ln \frac{r_f}{r_0} \,.
\label{bbb}
\eeq

The first   term in  (\ref{velvet}) gives,
in the first approximation, the wall tension $T_{\rm w}$ times the
area $\pi r_f^2$. The second term
can be written as  follows:
\beq
 \int d^3 x \, \xi \, \frac{F^{*}_3}{2}=\frac{\xi}{2} \left( \int dx_1 dx_2 F^{*}_3
 \right) \, L \approx
2 \pi \, \xi \, L\,.
\label{gargiulo}
\eeq
reflecting the fact that the magnetic
flux through different sections in the
$x_1,x_2$ plane is conserved and equal to $4 \pi$.
The expression for $L$ is given in Eqs.~(\ref{mantwo}) and (\ref{bbb}).
The estimate above is good at order ${O}(\Delta m / g^2)$
because  the coordinate  $z$ corresponding to the
left edge has uncertainty of the order
of the wall thickness $R$:
the slices in the $x_1,x_2$ plane
with radii close to $r_f$ fail to capture all the magnetic flux.

\begin{figure}[h]
\epsfxsize=5cm
\centerline{\epsfbox{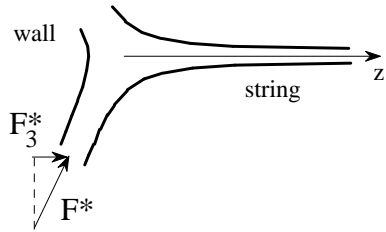}}
\caption{\footnotesize Direction of the magnetic flux inside the wall.}
\label{flux}
\end{figure}

The energy in  Eq. (\ref{gargiulo})
contains both the energy due to the ``actual" string tension (which
is $2 \pi \xi \ell $) and the energies $E^G +E^H$
(i.e. the surface bending plus the world-volume charge
energy, which is equal    $2 \pi \xi a$,  in total).
Now, we will derive this result from a  direct calculation
at small $z$. We have
\beq
  E_{0<z<a}=  \frac{\xi}{2}\int_0^a dz  \int dx_1 dx_2  F^{*}_3   =
 \frac{\xi}{2} \,2\pi\int rdr\,\int_{\zeta-R/2}^{\zeta+R/2}
dz|F^{*}||\partial_r\zeta |,
\label{stringe}
\eeq
where $F^{*}$ is the vector of the magnetic flux inside the wall,
while the factor  $\partial_r\zeta $ produces its $z$-component, see
Fig.~\ref{flux}. The magnetic flux inside the string is directed along
the $z$-axis and is equal  to $4\pi$. This flux is spread out inside the
wall and directed along the wall surface to the point of the wall-string
junction, see Fig.~\ref{flux}. We have inside the wall
\beq
|F^{*}| =\frac{2}{R}\,\frac1r,
\label{wflux}
\eeq
see \cite{sy1} for details. Substituting this into Eq. (\ref{stringe})
and using (\ref{shapeb}) we finally get
\beq
  E_{0<z<a}=  \frac{\xi}{2}\int_0^a dz  \int dx_1 dx_2  F^{*}_3 =
  \frac{2\pi\xi}{\Delta m}
\,\ln\frac{r_f}{r_0}=2 \pi \xi\,a.
\label{Estring}
\eeq

There is an ambiguity in defining the distance $L$ due
to the fact that the wall
has a finite thickness
of order $$\frac{ \Delta m}{g^2 \xi}\,.$$
This  is reflected
in an uncertainty in this estimate of the energy
of the order of
$${O}\left(\frac{\Delta m}{g^2}\right)\,.$$
The last term of (\ref{velvet})
gives a finite
negative contribution $$- \frac{4 \pi}{g^2} \Delta m$$
discussed in Ref. \cite{sakai-tong};
this is of the same order as the uncertainty
due to ambiguity in the   definition of
the integration domain of the Hamiltonian.
The total energy of the soliton,
which is defined up to  a quantity of the order of
$${O}\left(\frac{\Delta m}{g^2}\right)\,,$$
is given by the following expression:
\beqn
 E &=&
\int d^3 x \mathcal{H}\approx T_{\rm w} \pi r_f^2 + T_{\rm s} L
\nonumber\\[3mm]
&=&
    T_{\rm w} \pi r_f^2
    +\frac{2\pi\xi}{\Delta m}\,\ln\frac{r_f}{r_0}+ T_{\rm s} \ell\,.
\label{fly}
\eeqn

Let us introduce $S_0$,
the area of a would-be unbent wall,
\beq
S_0 =\pi r_f^2\,.
\eeq
Equation (\ref{fly})
implies that the logarithmic terms in
the energy of the boojum configuration   can be absorbed
in various ways. For instance,
in terms of  the actual area $S$ of the curved wall,
$$
S= S_0 + \Delta S
$$
and measuring the string length from the origin,
(i.e. length$ =\ell$ in Fig.~\ref{auone}) we have
\beqn
E
&=& T_{\rm w} S_0 +E^H+E^G+T_{\rm s} \ell
\nonumber\\[2mm]
&=&
T_{\rm w} S+E^G+T_{\rm s} \ell
,
\label{teone}
\eeqn
where $E^G$ is calculated in Eq. (\ref{fbc}).

On the other hand, in terms of an imaginary string length
$L= a+\ell$ the total energy is given by
\beq
E=T_{\rm w} S_0 +T_{\rm s} L\,.
\label{tetwo}
\eeq
We see that the energy gain in the wall
world-volume equals to the energy loss due to contraction of $L$.
 We will use this fact in  Sect.~\ref{tvewi}.

In both cases there is  an intrinsic uncertainty in
defining the above geometric parameters.
It is clear that the  curved wall area $S$ can be defined
only up to $\sim r_0^2$ which leads to the
energy uncertainty
$$
\sim T_{\rm w}r_0^2 \sim \Delta m/g^2\,.
$$
If $S$ is fixed to be $\pi r_f^2$ (i.e. the bending is excluded)
and is well-defined,
the uncertainty in $L$ is of the order of $R$
implying the energy uncertainty
$$\sim T_{\rm s} R \sim \Delta m/g^2
\,.
$$
Both derivations perfectly match. The resulting uncertainty is
of the same order as the negative binding term of Sakai and Tong.

\section{String interaction in multi-boojums}
\label{tvewi}

\begin{figure}[h]
\begin{center}
\leavevmode
\epsfxsize 12  cm
\epsffile{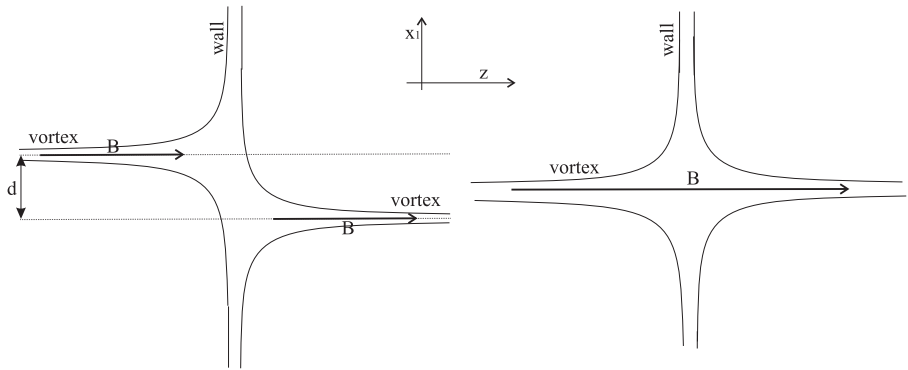}
\end{center}
\caption{\footnotesize Vortices  on opposite sides of the wall which
carry the same positive U$(1)$ flux in the $z$ direction. This
is a deformation with the same energy
of the configuration of a vortex crossing the wall.}
\label{bbking}
\end{figure}

\begin{figure}[h]
\begin{center}
\leavevmode
\epsfxsize 7  cm
\epsffile{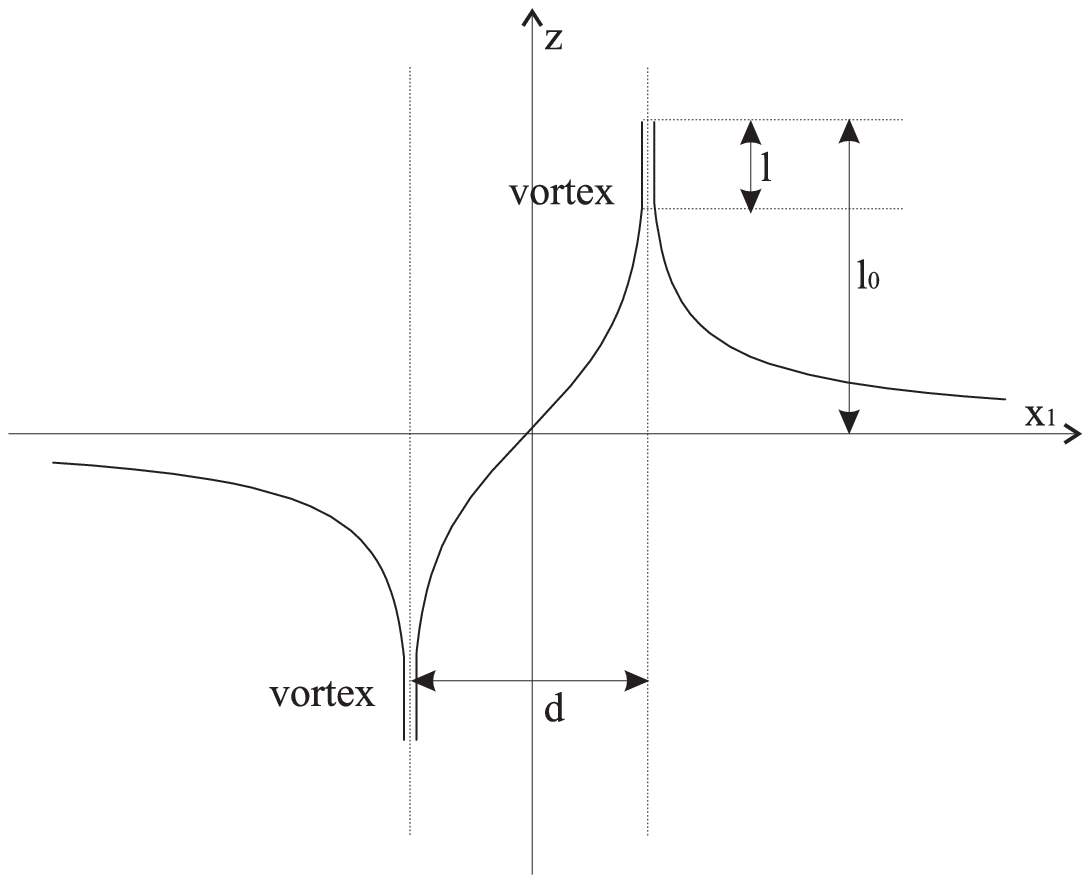}
\end{center}
\caption{\footnotesize Geometry of a two-boojums configuration
with  the vortices on the different sides of the wall
at large
distance $d$. The wall thickness is considered  zero in this picture.
The bending of the wall is given by $z=\frac{1}{\Delta m}\ln \left| \frac{x_1+d/2}{x_1-d/2} \right|
$.
 }
\label{fantomas}
\end{figure}

\begin{figure}[h]
\begin{center}
\leavevmode
\epsfxsize 10  cm
\epsffile{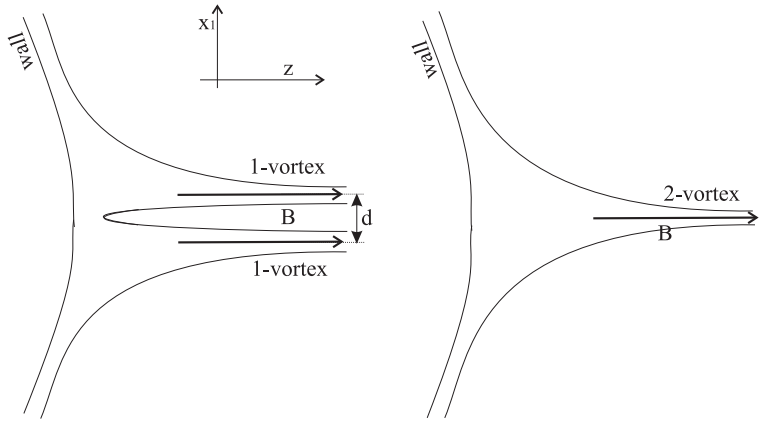}
\end{center}
\caption{\footnotesize Two elementary vortices  on the same side of the wall which
carry the same positive U$(1)$ flux in the $z$ direction. This
is a deformation of the configuration of a two-winding vortex ending on a wall.}
\label{clapton}
\end{figure}

\begin{figure}[h]
\begin{center}
\leavevmode
\epsfxsize 7  cm
\epsffile{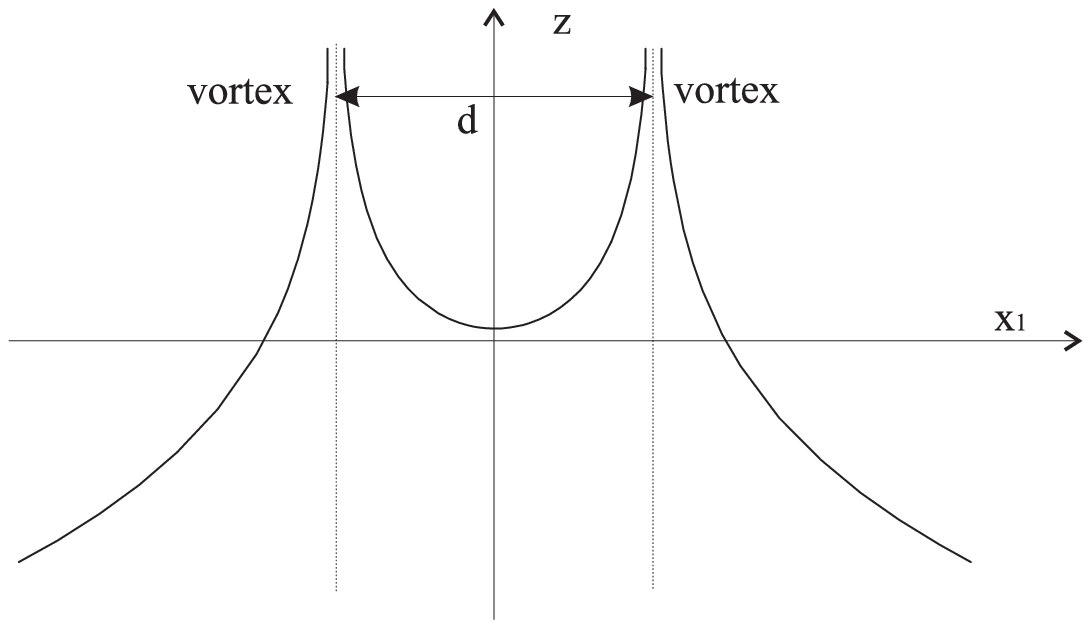}
\end{center}
\caption{\footnotesize Geometry of a two-boojums configuration
with both the vortices on the same side of the wall
 at large
distance $d$. The wall thickness is considered  zero in this picture.
The bending of the wall is given by
$z=\frac{1}{\Delta m}\ln \left| \frac{r_0^2}{(x_1-d/2)(x_1+d/2)} \right|$.
 }
\label{pulp}
\end{figure}

\begin{figure}[h]
\begin{center}
\leavevmode
\epsfxsize 10  cm
\epsffile{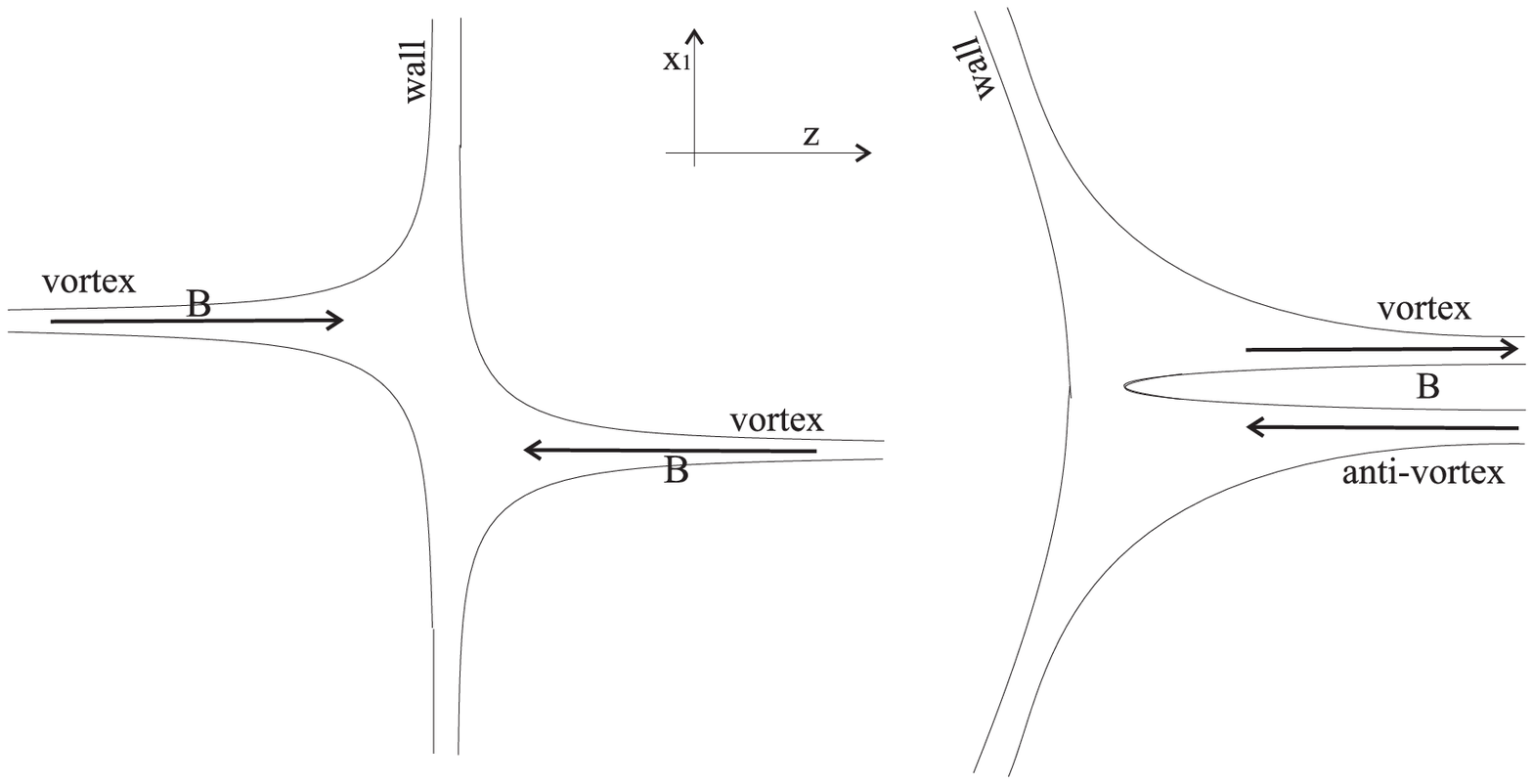}
\end{center}
\caption{\footnotesize No BPS solution exists for these configurations;
in the case on the left the force between the junctions is repulsive
while in the case on the right the force is attractive.}
\label{albertcollins}
\end{figure}

Let us first of all consider the situation in which
the two vortices are on opposite sides of the wall and
carry the same positive U$(1)$ flux in the $z$ direction
(see Fig. \ref{bbking}).
The distance between the two strings in the $x_1,x_2$
plane is denoted by $d$.
The two vortices pull the wall from different directions and
so there is no asymptotic bending at a distance much larger than $d$.
The two wall-vortex junctions
can interact throughout the gauge and   Higgs fields;
it is easy to calculate these interactions in the
wall world-volume theory in the regime of large $d$.

The gauge flux produced by two junctions has the form
\beq
F_{0i}^{2+1}=\frac{e^2_{2+1}}{2 \pi } \left[
\frac{(x-x_1)_i}{|x-x_1|^2}   -\frac{(x-x_2)_i}{|x-x_2|^2}  \right],
\label{2juncfl}
\eeq
where $i=1,2$ and $x_1$ and $x_2$ are the positions of two junctions on the
wall, $d=x_1-x_2$.
 Note the opposite signs of two terms here. This is because there
 is no gauge flux inside the wall at large $r$
in the configuration shown in Fig.~\ref{bbking}. In other words the
electric charges of two junctions are opposite. Substituting
this flux in the world volume action (\ref{3wwa}) we get
the energy of two junctions induced by the gauge interaction
\beq
V^G=\frac{2 \pi\xi}{\Delta m} \ln \frac{d}{r_0}.
\label{gpot}
\eeq
This potential gives an attractive force. Note the cancellation
of IR-divergent ``self-energy" part. This happens because there is  no
flux at infinity in this configuration.

Now consider the potential  induced by Higgs  interactions.
The wall bending produced by two junctions has the  form
(see Fig. \ref{fantomas}):
\beq
\partial_i \zeta= \frac1{\Delta m} \frac{(x-x_1)_i}{|x-x_1|^2}
- \frac1{\Delta m} \frac{(x-x_2)_i}{|x-x_2|^2}.
\label{2juncb}
\eeq
Two terms here have different signs again because there is no
wall bending at large $r$ for the configuration in Fig.~\ref{bbking}.
Substituting this in (\ref{3wwa})   we get
\beq
V^H=\frac{2 \pi\xi}{\Delta m} \ln \frac{d}{r_0}.
\label{hpot}
\eeq
This potential is also attractive. Note again the cancellation of
the IR-divergent part
(note that
it is possible to provide a bulk interpretation to this contribution as due to a
stretch of one flux tube in the presence of the wall logarithmic
deformation caused by the other).

So far we considered interaction induced by the wall bending and the
presence of the flux inside the wall. There is also another effect
which we have to  take into account: the length of both string changes
as we change the distance $d$ between junctions. Let us denote by
$l_0$ the length of each string at $d=0$ when the wall is not bent.
Then the actual length of the string as a function of distance $d$ is
\beq
l=l_0-\frac1{\Delta m} \ln \frac{d}{r_0},
\label{strlength}
\eeq
see (\ref{shapeb}), (\ref{2juncb}) and Fig. \ref{fantomas}.
 Note that ambiguity in defining
the distance between the junctions results in the ambiguity in the argument
of the logarithm. The coefficient in front of the logarithm is determined
unambiguously. Taking into account that the string tension is
equal  to $2\pi\xi$ and we have two strings we get the
following repulsive  potential
\beq
V^S=-\frac{4 \pi\xi}{\Delta m} \ln \frac{d}{r_0}.
\label{spot}
\eeq
Assembling  all three contributions in (\ref{gpot}),  (\ref{hpot})
and (\ref{spot}) we see that string junctions do not interact,
\beq
V=V^G +V^H +V^S =0.
\label{jpot1}
\eeq
Thus, there is no effective force in the wall world-volume between the two
junctions.

This calculation suggests that there is no force between the
two vortices not only in the logarithmic regime, but
also in the small-distance regime (where the logarithmic approximation
that we exploit
fails). We expect   that  $d$
is a modulus of the composite soliton. It should be possible then to
realize this configuration for arbitrary $d$ as an 1/4-BPS
solution of equations (\ref{bpseq}).

This type of solution allows us to describe the deformation of
an ANO vortex traversing the domain wall as two vortices ending
on the wall at separate points.
The situation is  similar to what happens in string theory
to a string traversing a brane (see, for example, \cite{callan-maldacena} for
a discussion of the problem using the Born-Infeld action).

The total energy of this composite BPS object is given by the sum of the three central
charges (see Eq. (\ref{velvet})).
Let us consider the energy inside a large  cylinder
$$x_1^2+x_2^2<r_f^2$$
of length $l$ in the $z$ direction;
this domain is chosen in such a way that it contains
both the vortex junctions and  both   flat faces of the
cylinder are at large distances from the wall plane.
The result for the energy is
\beq \int d^3 x \mathcal{H}= T_{\rm w} \pi r_f^2 + T_{\rm s} l
- \frac{8 \pi}{g^2}\,  \Delta m\, ,
\label{tcont}
\eeq
$(l=2l_0)$. The constant term
due to the Lorentz-scalar central charge
(which is discussed in Ref. \cite{sakai-tong}) is {\em well-defined in this
expression}. Note that ambiguity in the arguments of the logarithms in
Eqs.~(\ref{gpot}), (\ref{hpot}) and (\ref{spot}) produces an effect of the order
of $\xi/\Delta m$, which is much smaller than the last term in the right-hand
side of Eq.~(\ref{tcont}).

Our theory is  Abelian  and, as such,   does not
have  monopole solutions. However, if we started from a non-Abelian
$\mathcal{N}=2$ theory with the gauge group $U(2)$,
we would have confined monopoles attached to the end points of the
flux tubes,  both in the first and  the second vacua
(see Refs.~{\cite{tong-monopolo,SY-vortici}}).
The constant term in Eq.~(\ref{tcont}) can be
interpreted then as the mass difference between these two
monopoles (see also Ref. \cite{sakai-tong}
for a qualitative description of
a monopole passing through a domain
wall with two collinear strings on each side).

There is another configuration in which the world-volume
Coulomb and Higgs repulsion are exactly canceled
by the  ``string length" attraction:
the one in which the two vortices are both on the same side
of the wall with the same magnetic flux orientation, see
Figs.~\ref{clapton} and \ref{pulp}.
This is consistent with the behavior of the
vortices in the bulk:  far from the wall there is no force between them
because they are relatively BPS.

In this case, as previously, we expect the composite
soliton to be $1/4$ BPS saturated, and the relative distance
$d$ to  be a modulus.
The computation of energy of this system is completely
similar to the one for a single string ending on a wall;
in particular, there is a $\ln r$ divergence in the binding energy
due to the $(1/2,1/2)$
central charge.

On the other hand, the situation is different
if the vortices have the opposite magnetic fluxes in the $z$ direction,
see Fig.~\ref{albertcollins}.
If the vortices stay on different sides of the wall,
 the Higgs interaction between their junctions
is attractive while  the electric
interaction  is repulsive. These effects cancel each other.
The resulting total potential is due to the ``string length" repulsion,
\beq
V=-\frac{4 \pi\xi}{\Delta m} \ln \frac{d}{r_0}\,.
\label{jpot2}
\eeq
There is no chance to obtain a BPS configuration.

A similar thing happens when   two vortices are on the same side
of the wall:
the total potential is due to the ``string length" attraction,
\beq
V=\frac{4 \pi\xi}{\Delta m} \ln \frac{d}{r_0}\,.
\label{jpot3}
\eeq
The vortex and the anti-vortex will attract and annihilate each other;
the interaction will be stronger on the wall world-volume where the
theory  is
in the Coulomb phase, and will  be exponentially suppressed  in the bulk by
distance between the vortices.

\section{Boojum self-energy: the second try\,\protect\footnote{
This section was written as a result of discussions with   
David Tong to whom we are deeply grateful.}}
\label{bsest}

In Sect.~\ref{vwj} we
calculated the self-energy of a single boojum.
We found   this energy to be  logarithmically divergent
in   the IR limit  $r_f\to\infty$, see Eq. (\ref{eto}).
The attempt to isolate a finite binding energy of the boojum
of the order of $\Delta m/g^2$ failed because of a
definition-dependence of the parameters $\ell$ and $S$.

On the other hand, in Sect.~\ref{tvewi} it was explained
that in a certain two-boojum configuration all logarithmic terms
cancel, see Eq.~(\ref{jpot1}), relevant geometric parameters
$l$ and $S=\pi r_f^2$ become well-defined
and, as a result,  the two-boojum energy contains a well-defined
\cite{sakai-tong} finite
term $-8\pi\Delta m/g^2$, see Eq.~(\ref{tcont}).

The question arises as to whether one can take advantage
of this situation
to unambiguously determine the single-boojum binding energy.
To this end one can try to use the following natural strategy.

\begin{figure}[h]
\begin{center}
\leavevmode
\epsfxsize 8  cm
\epsffile{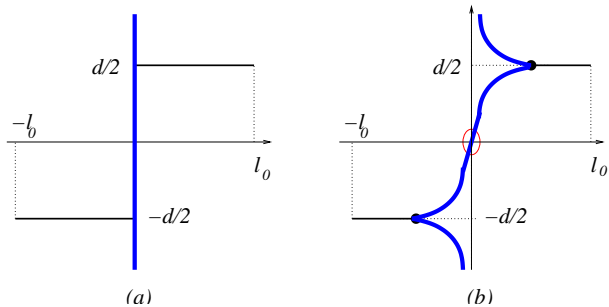}
\end{center}
\caption{\footnotesize A hypothetical (a) and actual (b) boojum-antiboojum
configuration. In the logarithmic approximation both have the same energies.
Since the {\em bona fide} solution is presented by (b),
it must have a negative finite binding energy.}
\label{twog}
\end{figure}

Consider the two-boojum configuration depicted in
Fig.~\ref{bbking}. We will refer to this configuration as
to boojum-antiboojum system because
it is characterized by the absence of both, the wall bending and flux at
$r\to\infty$ with $d$ fixed. As was mentioned, all logarithms cancel,
but a finite negative binding energy {\em must} exists, as is obvious {\em a priori}
from comparison of two graphs presented in Fig.~\ref{twog}.

\begin{figure}[h]
\begin{center}
\leavevmode
\epsfxsize 6 cm
\epsffile{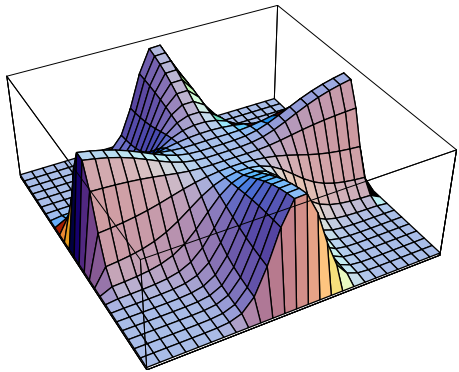}
\end{center}
\caption{\footnotesize Energy density of the domain wall junction}
\label{led}
\end{figure}

The very fact of negative binding energies for junctions is not surprising and
is wide-spread. For instance, a negative tension for the multi-domain wall
junction in a Wess-Zumino model was found in Ref.~\cite{Oda}.
The reason underlying this phenomenon was explained in \cite{Tonis}:
the local energy density is diluted inside the junction.
Figure~\ref{led} borrowed from Ref.~\cite{Tonis} illustrates this statement.
It clearly shows that the energy density inside the junction is lower than that
inside the walls (why this is the case is also understood).

Let us
turn to the boojum-antiboojum system and start from vanishing separation, $d=0$
see Fig.~\ref{bbking}. The plot on the right corresponds to the string crossing the wall with no long range interactions.
There is no doubt that the binding energy
of this configuration, $-8\pi\Delta m/g^2$, is localized inside the bulge at the origin,
where the energy density is diluted compared to that outside the bulge.
(It is not difficult to see that the energy density inside the wall and both flux tubes
is of the order of $g^2\xi^2$.)

Now, let us gradually increase $d$ passing to   the IR limit
$d\gg R, \,r_0$. There are two logical possibilities.
(i) The overall binding energy $-8\pi\Delta m/g^2$,
which remains intact, splits into two domains of energy-density deficit,
each localized at the position of the corresponding junction.\,\footnote{This option is advocated by David Tong.}
Then, this would provide us with a sensible definition/determination of
a finite binding energy of the isolated boojum in the form
\beq
E=-\frac{8\pi}{g^2}\,\Delta m\, \cdot \left(\frac{1}{2}\right)
\label{booe}
\eeq
per boojum, in full accordance with
Ref.~\cite{sakai-tong}. (ii) The total binding energy $-8\pi\Delta m/g^2$
is localized near the junctions {\em and elsewhere}.
On physical grounds, it cannot spread evenly between the junctions.
We would have captured such a spread in the logarithmic terms.
On the other hand, it is  intuitively clear that the {\em third localized
domain} of energy-density deficit might emerge near the origin (see the red
oval in Fig.~\ref{twog}$b$). This is a transition domain between
the positive and negative logarithmic bending;
in this domain the wall is bent to a lesser extent than outside this domain.
Thus, the energy due to the Higgs field is expected to be lower here
than outside this domain. On the other hand, the energy
due to the gauge field flux may or may not be higher.
One can estimate the order of magnitude of possible 
energy deficit near the origin. Using the fact that 
the local energy density $\sim g^2\xi^2$, in conjunction
with Eqs. (\ref{til}) and (\ref{tt}),
we arrive at the energy deficit $\sim \Delta m/g^2$, i.e. of the same order 
of magnitude as in Eq. (\ref{booe}).
If so, then the binding energy of the isolated boojum
cannot be defined in this way, i.e. as  half of the binding energy of
the boojum-antiboojum configuration at large $d$.

Needless to say, the above intuitive observation of a central domain
of an energy deficit can only be firmly confirmed  
by a direct numerical
solution for the boojum-antiboojum configuration
analogous to that carried out in Ref.~\cite{Tonis}.
Then we could analyze the local energy density
point-by-point inside the boojum-antiboojum configuration
and observe (or rule out) an energy-density deficit in the central domain.

\section{Conclusions}
\label{conclu}

In the present paper we studied 1/4-BPS boojums in the simplest possible
model,  \ntwo QED. We calculated the energy of an isolated boojum and found that
it is logarithmically divergent in the infrared domain. This divergence appears
because the string end  represents an electric charge in the effective
theory on the wall, and moreover, the string produces a
logarithmic bending of the wall. We
compute this energy in three different ways.  First, we calculate it
from the point of view of (2+1)-dimensional effective theory on the
wall. Then we rederive the same result in the bulk theory: it comes
from an effective increase of the wall surface due to its bending and
the string magnetic flux spread inside the wall. Finally we confirm
this result studying  contributions to the central charges (\ref{velvet})
of different BPS objects.

There are uncertainties in defining the area of the wall and the length
of the string. These uncertainties produce ambiguous finite
contributions to the boojum energy, of the order of
the monopole mass  $\sim m/g^2$. They are of the same order
as the result for the boojum energy obtained in Ref. \cite{sakai-tong}.

Next we study multi-boojums.
Interaction potential of two strings ending on the
same wall is considered. In the bulk these interactions are
exponentially suppressed,
while on the wall they are of a long range type due to the presence
of massless fields in the world-volume theory on the wall.
We find two configurations (see Fig.~\ref{bbking} and
Fig.~\ref{clapton}) for which the interaction potential disappears.
They should correspond to 1/4-BPS solutions of the first-order equations
(\ref{bpseq}).

Since the self-energy of the isolated boojum has a logarithmic part
which makes it virtually impossible to detect a finite
contribution discussed in   \cite{sakai-tong}
directly from the single-boojum configuration,
we make an attempt to define it from the  boojum-antiboojum configuration
whose energy has no logarithmic parts. The outcome remains inconclusive,
depending on which of the two conjectures formulated in Sect.~\ref{bsest} wins.
The winner can be determined by a direct numerical calculation.

\section*{Acknowledgments}

We are grateful to Jarah Evslin for useful discussion and David Tong
for valuable communications and comments.

The work  of R.A. and M.S.  was
supported in part by DOE grant DE-FG02-94ER408.
The work of A.Y. was supported in part by Theoretical Physics Institute
at the University of Minnesota.

\end{document}